\documentclass[12pt]{article}
\usepackage{latexsym,amssymb,amsmath}
\newtheorem{theorem}{Theorem}[section]
\newtheorem{examp}{Example}[section]
\newtheorem{examps}{Examples}[section]

\newtheorem{lemma}[theorem]{Lemma}
\newtheorem{remark}{Remark}[section]
\newtheorem{remarks}[remark]{Remarks}
\newtheorem{proposition}[theorem]{Proposition}
\newtheorem{definition}{Definition}[section]

\def\begindef{\begin{definition}}
\def\enddef{\end{definition}}
\def\brem{\begin{remark}\rm}
\def\erem{\end{remark}}
\def\brems{\begin{remarks}\rm}
\def\erems{\end{remarks}}
\def\bexample{\begin{examp}\rm}
\def\eexample{\par\noindent $\square$ \end{examp}}
\def\bexamples{\begin{examps}\rm}
\def\eexamples{\par\noindent $\square$ \end{examps}}
\def\proof{\begingroup\noindent\bf Proof.\ \endgroup}
\def\endpf{\hfill $\square$\\\medskip\\}

\def\rref#1{(\ref{#1})}
\def\de{\partial}

\newcommand{\spec}[1]{{\rm Spec}(#1)}
\newcommand{\Span}[2]{{\rm span}_{#1}\{{#2}\}}
\newcommand{\Gr}{{\rm SGr}_\Lambda}
\newcommand{\Div}{{\rm Div}}
\newcommand{\pic}{{\rm Pic}}
\newcommand{\jac}{{\rm Jac}}
\newcommand{\Hom}{{\rm Hom}}
\newcommand{\rd}{{\rm rd}}
\newcommand{\spt}{{\rm sp}}
\newcommand{\sbin}[2]{\genfrac{[}{]}{0pt}{}{#1}{#2}}
\newcommand{\col}[2]{\genfrac{}{}{0pt}{}{#1}{#2}}
\newcommand{\di}{\mbox{d}}
\newcommand{\CA}{{\cal A}}

\newcommand{\CC}{{\cal C}}
\newcommand{\uCC}{{\underline{\cal C}}}
\newcommand{\tCC}{{\tilde{\cal C}}}
\newcommand{\CD}{{\cal D}}
\newcommand{\CE}{{\cal E}}

\newcommand{\CJ}{{\cal J}}

\newcommand{\CL}{{\cal L}}

\newcommand{\CN}{{\cal N}}
\newcommand{\CO}{{\cal O}}

\newcommand{\CQ}{{\cal Q}}

\newcommand{\CS}{{\cal S}}

\newcommand{\CU}{{\cal U}}
\newcommand{\CV}{{\cal V}}

\newcommand{\CX}{{\cal X}}

\newcommand{\CZ}{{\cal Z}}
\makeatletter
\@addtoreset{equation}{section}

\makeatother
\newcommand{\BC}{{\mathbb C}}

\newcommand{\NN}{{\mathbb N}}

\def\la{\lambda}        \def\La{\Lambda}
                 \def\tDe{\tilde{\Delta}}

\def\Fdb{{Fa\`a di Bruno}}

\newcommand{\dpt}[2]{{\displaystyle{\frac{\partial #1}{\partial t_{#2}}}}}
\newcommand{\hh}{\hat{h}}

\newcommand{\Ha}[1]{H^{(#1)}}
\newcommand{\Da}[1]{\Delta^{(#1)}}
\newcommand{\tDa}[1]{\tilde{\Delta}^{(#1)}}
\newcommand{\SH}[1]{\wid{H}^{(#1)}}

\newcommand{\sh}[1]{\hat{h}^{(#1)}}

\def\ger{hierarch}

\def\ger{hierarch}

\newcommand{\wid}[1]{\widehat{#1}}
\begin{document}
%\baselineskip=14pt
%\begin{titlepage}
\begin{flushright}
Ref. SISSA 36/2000/FM
\end{flushright}
\vspace{1.truecm}
\begin{center}
{\huge A note on the super Krichever map}
\end{center}
\vspace{1,2truecm}
\makeatletter
\begin{center}
{\large
Gregorio Falqui, Cesare Reina and Alessandro Zampa\\ \bigskip
SISSA, Via Beirut 2/4, I-34014 Trieste, Italy}\\
E--mail: falqui@sissa.it, reina@sissa.it,
zampa@fm.sissa.it\\
\vspace{.3truecm}
%\today
\end{center}
\makeatother
\vspace{1.5truecm}
\noindent {\bf Abstract}. 
We consider the geometrical aspects of the Krichever map in the context of
Jacobian Super KP \ger y. We use the representation of the \ger y based on the 
\Fdb\ recursion relations, considered as
the cocycle condition for the natural double complex associated with 
the deformations of super Krichever data. Our approach is based on  the 
construction of the universal super divisor (of degree $g$), 
and a local universal family of geometric data which give the map into the 
Super Grassmannian.
\vspace{.5truecm}
%\setcounter{footnote}{0}

%%%%%%%%%%%%%%%%%%%%%%%%%%%%%%%%%%%%%%%%%%%%%%%%%%%%%%%%%%%%%%%%%%%%%%
%
%                      I N T R O D U Z I O N E 
%
%%%%%%%%%%%%%%%%%%%%%%%%%%%%%%%%%%%%%%%%%%%%%%%%%%%%%%%%%%%%%%%%%%%%%%

\section{Introduction}\label{setc:0}
In this note we study the geometrical setting of the super Krichever map in
analogy to the standard non graded case~\cite{FRZ1}. This map is an essential
tool in the analysis of the algebraic geometric solutions to integrable systems
of soliton type (see, e.g.,~\cite{MuAlg}). Its super extension has already been
introduced in~\cite{MuRab}, and studied in~\cite{Rab,BerRab}. The essential
difference in our approach is to make full profit of the so--called \Fdb\ 
approach to the KP theory~\cite{FMP} and its super 
generalization~\cite{FRZ2}, where the 
equivalence of this approach to the standard
differential-operator approach to the Jacobian SKP~\cite{Mu,Rab} is
proved. It turns out, as in the classical case, that the \Fdb\ recursion
relation is (related to) the first  cocycle condition for the hypercohomology
group which controls the infinitesimal deformations of the spectral super line
bundle together with its meromorphic sections.

The basic advantage of this approach is that it is directly related to the
(Super)Grassmannian description of the \ger y, and has an intrinsic geometrical
meaning. In particular, we can avoid the difficult initial step of the
introduction of the Baker--Akhiezer function, go on with the natural
development of the geometrical construction, and recover the existence of the
BA function at the end.  

As in the classical case, the important technical tool is to construct a local
universal deformation of the initial super line bundle, since the cocycle
condition comes by considering point-wise the germ of such a deformation as a
vector field on the base. This is a difficult point because we lack a
sound definition of the Super Jacobian of a super curve $\CC$. 
Indeed looking
at the transition functions, one would say that this is the cohomology group
$H^1(\CC,\CO^{\times}_{0})$, where $\CO^{\times}_{0}$ is the 
sheaf of units in the even part of the structure sheaf. Unfortunately, this 
set-up is not fully satisfactory because there are no naturally 
defined odd deformation directions.
The way out we present in this paper is to work with the moduli space
$S_g\tilde\CC$ of
effective superdivisors of degree $g$ or, which is the same, with the
$g$--fold symmetric product~\cite{DPHRSS} of the dual curve~\cite{DRS,BerRab}.
This is a supervariety with enough odd parameters over which we have a
universal effective divisor, and we expect that, as in the classical case,
the ``Super Jacobian'' will appear as a quotient of  $S_g\tilde\CC$. 

The scheme of the paper is as follows: in Section~\ref{sect:1} we briefly
recall the \Fdb\ recursion relations, their connection
with the JSKP \ger y, and with the
Krichever map, referring to~\cite{FRZ2} for more details. In 
Section~\ref{sect:2} we give a brief {\em resum\`e} of the tools from 
deformation theory needed in the sequel. In Section~\ref{sect:3} we construct
the symmetric powers of the (dual) supercurve as a supervariety, and we prove
the existence of a universal superdivisor. In the last Section we exploit the
cohomological meaning of the \Fdb\ recursion relations to insure that it gives
a flow on the space of super Krichever data and, through the super Krichever
map, the JSKP flow on the algebraic geometrical loci in the Super Grassmannian.
Finally, in Appendix A we recall some basic definitions of the theory of
super curves used in the paper.    

%%%%%%%%%%%%%%%%%%%%%%%%%%%%%%%%%%%%%%%%%%%%%%%%%%%%%%%%%%%%%%%%%%%%%%
%
%                             J S K P
%
%%%%%%%%%%%%%%%%%%%%%%%%%%%%%%%%%%%%%%%%%%%%%%%%%%%%%%%%%%%%%%%%%%%%%%

\section{The Jacobian super KP hierarchy}\label{sect:1}

Let us start by fixing some notations.
We denote by $\La$ a generic Grassmann algebra over $\BC$,
$B:=\La[[x,\varphi]]$
is the $\La$--algebra of formal power series in the variables $x$ (even) and
$\varphi$ (odd) and $\CD:=\de_\varphi+\varphi\de_x$.
The ring of formal super pseudo--differential operators over $X:=\spec{B}$
is the space of formal series
$$L:=\sum_{j\ge 0}u_j\CD^{n-j},\qquad u_j\in B$$
endowed with the product induced by the super Leibniz rule
$$\CD^k\cdot f=
\sum_{j\ge 0}(-1)^{\bar{f}(k-j)}\sbin{k}{j}f^{(j)}\CD^{k-j},$$
where $\bar{f}$ denotes the parity of $f$,
$f^{(j)}=\CD^j(f)$ and $\sbin{k}{j}$
is the super binomial coefficient \cite{MaRad}.

Mulase and Rabin defined the Jacobian super KP hierarchy \cite{Mu,Rab} as the
following set of evolutionary equations for the even dressing operator
$S:=1+\sum_{j>0}s_j\CD^{-j}$:
$$\left\{\begin{array}{l}
\de_{t_{2k}}S:=-(S\CD^{2k}S^{-1})_-S=-(S\de_x^kS^{-1})_-S \\
\\
\de_{t_{2k-1}}S:=-(S(\CD^{2k-1}-\varphi\CD^{2k})S^{-1})_-S=
                 -(S\de_\varphi\de_x^{k-1}S^{-1})_-S,
\end{array}\right.$$
where $L_-$ is the pure pseudo--differential part of $L$ and the time 
$t_k$ has parity $k\bmod 2$.
One of the features which distinguishes this hierarchy from that of Manin and
Radul \cite{MaRad} is that for algebraic geometric solutions the equations
describe super--commuting linear flows on the super Jacobian of a super curve.
One way to approach this issue is to consider another description of the
hierarchy, using the super Fa\`a di Bruno polynomials instead of super
pseudo--differential operators.
We refer to \cite{FRZ2} for a detailed account and we only sketch
what is relevant to the present discussion.
Let $V:=\La((z^{-1}))\oplus\La((z^{-1}))\cdot\theta$ be the algebra of formal
Laurent series in the even variable $z^{-1}$ and odd variable $\theta$,
let $V_+:=\La[z,\theta]$, $V_-:=\La[[z^{-1},\theta]]\cdot z^{-1}$ and
let $V_B:=V\otimes_\La B$.
The basic object of this formulation is the odd Fa\`a di Bruno generator
$$\hh(z,\theta;x,\varphi):=\theta+\varphi z+O(z^{-1})\in V_B$$
where, abusing notations, we write $O(z^{-1})$ for an element of
$V_-\otimes_\La B$.
Out of $\hh$ we construct iteratively the Fa\`a di Bruno polynomials by
\begin{equation}\label{eq:sfdb}
\left\{\begin{array}{l}
\sh{0}:=1 \\
\\
\sh{k+1}:=(\CD+\hh)\sh{k}\qquad k\in\NN
\end{array}\right.
\end{equation}
and set $W_B:=\Span{B}{\sh{k}:\,k\in\NN}$.
It is then easy to show that there exists a unique basis
$\{\SH{k},\,k\in\NN\}$ of $W_B$, whose elements (called
``super currents'') have the form
\begin{equation}\label{eq:as}
\SH{2k+p}=\theta^pz^k+O(z^{-1})
\end{equation}
with $p=0,1$, in terms of which the equations of the Jacobian super
KP hierarchy become
\begin{equation}\label{eq:hskp}
\dpt{\hh}{k}=(-1)^k{\cal D}\SH{k}.
\end{equation}
Since $\SH{2}=\sh{2}$ we have $\de_{t_2}\equiv\de_x$.

The study of these equations finds its most appropriate 
and natural setting in the concept
of super universal Grassmannian $\Gr$ defined as follows \cite{Schw,BerRab}.
The filtration
$\cdots\subset V_{j-1}\subset V_j\subset V_{j+1}\subset\cdots\subset V$,
where $V_j=z^{j+1}V_-$, makes $V$ and its $\La$--submodule $V_+$ complete
topological spaces and $\Gr:=\Gr(V,V_+)$ is the set of closed free
$\La$--submodules $W$ of $V$ which are compatible with $V_+$ in the sense
that the restriction $\pi_W$ of the projection $\pi:V\to V_+$ to $W$ is a
Fredholm operator, i.e. its kernel (respectively cokernel) is a
$\La$--submodule (respectively a quotient $\La$--module)
of a finite rank free $\La$--module.
As in the commutative setting, $\Gr$ is the disjoint union of the
denumerable set of its components $\Gr^{(i)}$ labelled by the index $i_W$
of $\pi_W$, moreover each component acquires a structure of super scheme
by means of projective limits.
By definition, the space $W_B$ spanned by the $\SH{k}$'s gives rise to
a moving point of $\Gr$ and the super currents evolve
under JSKP along the equations of a dynamical system, the {\em super
central system} \cite{FRZ2}, which gives vector fields on the Grassmannian.
In particular one has that
\begin{equation}\label{eq:scomm}
\de_{t_j}\SH{k}=(-1)^{jk}\de_{t_k}\SH{j}.
\end{equation}

Whichever approach one takes, the link with algebraic geometric 
solutions is provided by the super Krichever
map \cite{MuRab} which associates a point $W$ of $\Gr$ to the datum
$(\CC,D,(z^{-1},\theta),\CL,\eta)$ of
\begin{itemize}
\item[i.] a $\La$--super--curve $\CC=(C,\CO_\CC)$ (see Appendix \ref{app:1}),
\item[ii.] an irreducible divisor $D$ on $\CC$ whose reduced support is a
smooth point $p_\infty\in C$,
\item[iii.] local coordinates $z^{-1}$ and $\theta$ in a neighbourhood
$U_0\ni p_\infty$,
\item[iv.] an invertible sheaf $\CL$ on $\CC$ and
\item[v.] a local trivialization $\eta$ of $\CL$ over $U_0$.
\end{itemize}
Let $\CL(\infty D)=\lim_{n\to\infty}\CL(nD)$ be the sheaf of sections of
$\CL$ with at most an arbitrary pole at $D$, then
$W=\eta(H^{0}(\CC,\CL(\infty D)))$.
Bergvelt and Rabin have shown in \cite{BerRab} that the $\La$--module
$H^{0}(\CC,\CL(\infty D))$ is free, so $W$ belongs indeed to $\Gr$.
We can invert the Krichever map on its image as explained in \cite{MuRab};
in particular, the ring of functions of $\CC$ which are holomorphic on
the open subset $U_1:=C-\{p_\infty\}$ is the subalgebra
$\CA_W\subset V$ of functions $f$ such that $f\cdot W\subset W$. $\CA_W$
is obviously graded.

As a consequence of equations \rref{eq:sfdb} and \rref{eq:hskp}, we recover
the same picture in our approach: 
\begin{proposition}[Isospectrality]\label{lem:isosp}
Let $\hh(x,\varphi,{\mathbf t})$ be a solution of the Jacobian super KP
hierarchy and denote by $W_T$ the space generated by the corresponding super
currents $\SH{k}(x,\varphi,{\mathbf t})$.
For any specialization $(x_0,\varphi_0,{\mathbf t}_0)$ of
$(x,\varphi,{\mathbf t})$ let
$\CA_{(x_0,\varphi_0,{\mathbf t}_0)}\subset V$ be the
$\La$--algebra of functions that map by multiplication $W_{T_0}$ into itself.
Then $\CA_{(x_0,\varphi_0,{\mathbf t}_0)}$ does not depend on
$(x_0,\varphi_0,{\mathbf t}_0)$.
\hfill$\square$
\end{proposition}
We limit ourselves to sketch the proof. We have to show that if $f\in
\CA_{(x_0,\varphi_0,{\mathbf t}_0)}$, then $fW_T\subset W_T$. 
Since $1\in W_{T}$, this is equivalent to showing that such an  
$f$ is in  $W_T$, because $f$ supercommutes with $\CD+\hat{h}$. 
Since $1\in W_{T_0}$ as well,
we can write
\[
f=\sum c_j\widetilde{\Ha{j}}
\]
where $\widetilde{\Ha{j}}$ denote the specialization of $\SH{j}$ at
$\mathbf{t}=\mathbf{t}_0$. We have to prove that, calling 
\[
f^\prime=\sum c_j\SH{j},
\] 
actually $f^\prime=f$,  that is that $\sum c_j\SH{j}$ is independent 
of the times $t_k$.

The identity:
$$f\cdot(\CD+\hat{h})^k=\sum_{j\ge 0}(-1)^{{{j(j+1)}\over{2}}+
k\bar{f}}{k\atopwithdelims[]j}(\CD+\hat{h})^{k-j}f^{(j)}$$
shows that
$\CD^kf^\prime\in W_T, \,\forall k$, so that 
$\CD f^\prime=0$. Similarly, one
proves that $\de_{t_k} f^\prime=0\,\forall k$.

%%%%%%%%%%%%%%%%%%%%%%%%%%%%%%%%%%%%%%%%%%%%%%%%%%%%%%%%%%%%%%%%%%%%%%
%
%               D E F O R M A T I O N   T H E O R Y
%
%%%%%%%%%%%%%%%%%%%%%%%%%%%%%%%%%%%%%%%%%%%%%%%%%%%%%%%%%%%%%%%%%%%%%%

\section{Deformation of super line bundles and of their sections}\label{sect:2}

The meaning of the Isospectrality Lemma \ref{lem:isosp} is that when the
solution is of algebraic geometric type the super curve $\CC$ (also called
the {\em spectral curve}) remains unaffected by the flows of the hierarchy.
Indeed, it is also true that the divisor $D$ and the coordinates
$(z^{-1},\theta)$ do not change, so the motion involves
only the line bundle $\CL$ and its local trivialization $\eta$.
Since our aim is to interpret geometrically the equations \rref{eq:sfdb} and
\rref{eq:hskp}, which are of differential type, and the super Krichever map
is defined in terms of sections of a super line bundle $\CL$, we have to
understand how the sections change when we deform $\CL$.

\begindef
Let $\CC$ be a $\La$--super--curve, $\CL$ an invertible sheaf on $\CC$,
$s$ a global section of $\CL$ and $(\CX,x)$ a pointed $\La$--super--scheme.
An {\em $\CX$--family of invertible sheaves on $\CC$} is an invertible sheaf
$\CL_\CX$ over $\CC\times_{\spec{\La}}\CX$.
A {\em deformation of $(\CL,s)$} over the pointed super--scheme $(\CX,x)$
is a triple $(\CL_\CX,\sigma,\rho)$ where
\begin{itemize}
\item[i.] $\CL_\CX$ is an $\CX$-family of invertible sheaves on $\CC$,
\item[ii.] $\sigma$ is a global section of $\CL_\CX$ and
\item[iii.] $\rho$ is an isomorphism $\rho:\CL\to\iota^*\CL_\CX$,
where $\iota:\CC\hookrightarrow\CC\times_{\spec{\La}}\CX$ is the embedding 
identifying $\CC$ with $\CC\times_{\spec{\La}}\{x\}$, such that
$\iota^*\sigma=\rho s$.
\end{itemize}
Two deformations $(\CL_\CX,\sigma,\rho)$ and $(\CN_\CX,\tau,\xi)$ of
$(\CL,s)$ over $(\CX,x)$ are isomorphic if and only if there exists an
isomorphism of sheaves $\eta:\CL_\CX\to\CN_\CX$ compatible with $\rho$
and $\xi$ ($\xi=\iota^*(\eta)\circ\rho$) and such that $tau=\eta(\sigma)$.
The line bundle $\CL_\CX\vert_{\CC\times_{\spec{\La}}\{x\}}\simeq\CL$ is
sometimes called the {\em central fibre} of the deformation.
Finally, an {\em infinitesimal deformation of $(\CL,s)$} is a
deformation over the ``one--point'' $\La$--super-scheme
$$\CE:=\spec{\La[t,\varepsilon]/\langle t^2,t\varepsilon\rangle},$$ 
where $t$ is even and $\varepsilon$ is odd.
\enddef

Let $\{U_j\}_{j\in J}$ be a covering by open affine sub--super--schemes of
$\CC$ and denote by $U_{j_1,\ldots,j_k}$ the intersection
$\bigcap_{l=1}^kU_{j_l}$, by $\CO_{j_1,\ldots,j_k}$ the super--commutative
ring of sections of $\CO_\CC$ over $U_{j_1,\ldots,j_k}$ and by
$\CL_{j_1,\ldots,j_k}$ the $\CO_{j_1,\ldots,j_k}$--module of sections of
$\CL$ over $U_{j_1,\ldots,j_k}$.
Finally, define
$$\left\{\begin{array}{l}
\CO_{j_1,\ldots,j_k}[t,\varepsilon]:=
\CO_{j_1,\ldots,j_k}\otimes_\La\CO_\CE \\
\\
\CL_{j_1,\ldots,j_k}[t,\varepsilon]:=
\CL_{j_1,\ldots,j_k}\otimes_\La\CO_\CE \\
\\
U_{j_1,\ldots,j_k}[t,\varepsilon]:=
\spec{\CO_{j_1,\ldots,j_k}[t,\varepsilon]}=
U_{j_1,\ldots,j_k}\times_{\spec{\La}}\CE
\end{array}\right..$$
Then, $\{U_j[t,\varepsilon]\}_{j\in J}$ is an open affine covering of
$\CC\times_{\spec{\La}}\CE$ and the exact sequence of sheaves
$$\begin{array}{ccccccccc}
0 & \longrightarrow & \CO_j & \longrightarrow &
    \CO_j[t,\varepsilon]_{0}^\times & \longrightarrow &
    \CO_{j,0}^\times & \longrightarrow & 1 \\
 & & & & & & & & \\
 & & f & \mapsto & 1+tf_0+\varepsilon f_1 & & & &
\end{array},$$
where $f_0$ and $f_1$ are the even and odd components of $f$, yields a group
isomorphism $\pic(U_j[t,\varepsilon])\simeq\pic(U_j)$ due to the fact
that the $U_j$'s are Stein (see \cite{V}, 1.3.8).
Thus, if $\CL_\CE$ is an infinitesimal deformation of $\CL$ then
$$\CL_\CE\vert_{U_j[t,\varepsilon]}\simeq
\left(\CL\vert_{U_j}\right)[t,\varepsilon],$$
so it is described as the gluing of the last modules by means of a suitable
isomorphism
$$G_{jk}:\CL_{jk}[t,\varepsilon]{\buildrel\simeq\over\longrightarrow}
\CL_{jk}[t,\varepsilon],$$
which in turn is given by the transition matrix
$$G_{jk}=\left(\begin{array}{ccc}
g_{jk} & 0 & 0 \\
\delta_tg_{jk} & g_{jk} & 0 \\
\delta_\varepsilon g_{jk} & 0 & g_{jk}
\end{array}\right),$$
where we express an element $\sigma_j\in\CL_j[t,\varepsilon]$,
$\sigma_j=f_j+t\delta_tf_j+\varepsilon\delta_\varepsilon f_j$
as a column vector
$(f_j,\delta_tf_j,\delta_\varepsilon f_j)^t$,
$\delta_tg_{jk}\in\CO_{jk,0}$, $\delta_\varepsilon g_{jk}\in\CO_{jk,1}$
and $g_{jk}$ is the transition function of $\CL$.
The cocycle condition for $G_{jk}$ implies that
$\{g_{jk}^{-1}(\delta_tg_{jk}+\delta_\varepsilon g_{jk})\}_{jk}$
is a $1$-cocycle $c_1$ on $\CC$ with values in $\CO_\CC$.
Clearly, if we change $c_1$ by a coboundary we get an isomorphic
infinitesimal deformation of the invertible sheaf $\CL$.
Hence, the set of isomorphism classes of infinitesimal deformations of
$\CL$ is isomorphic to $H^1(\CC,\CO_\CC)$.
If we have a deformation $\CL_\CX$ of $\CL$ over $(\CX,x)$ and
$v:\CE\to\CX$ is a ''tangent vector`` to $\CX$ at $x$, then the pull--back of
$\CL_\CX$ under $id_\CC\times v$ is an infinitesimal deformation of
$\CL$ and corresponds by the above argument to a class
$[c_1]\in H^1(\CC,\CO_\CC)$.
This defines the Kodaira--Spencer map $KS:T_{x}\CX\to H^1(\CC,\CO_\CC)$  
of the deformation.

Now we consider the deformation
$\sigma\in H^0(\CC\times_{\spec{\La}}\CE,\CL_\CE)$ of $s\in H^0(\CC,\CL)$.
Let us write the local expression of $\sigma$ as above:
$\sigma_j=f_j+t\delta_tf_j+\varepsilon\delta_\varepsilon f_j$,
where $f_j$ is the local function representing $s$.
Then, the cocycle condition for $\sigma$ to be a global section reads
\begin{equation}\label{eq:def}
\left\{\begin{array}{l}
g_{jk}^{-1}\delta_tf_j=\delta_tf_k+g_{jk}^{-1}\delta_tg_{jk}f_k\\
\\
g_{jk}^{-1}\delta_\varepsilon f_j=\delta_\varepsilon f_k+
g_{jk}^{-1}\delta_\varepsilon g_{jk}f_k
\end{array}\right..
\end{equation}
The meaning of these two equations is the following (see e.g.
\cite{W} and the Appendix of \cite{FRZ1}): the triple
$(\{U_j\}_j,\{g_{jk}^{-1}(\delta_tg_{jk}+\delta_\varepsilon g_{jk})\}_{jk},
\{\delta_tf_j+\delta_\varepsilon f_j\}_j)$
gives rise to a class $\gamma_1\in{\mathbb H}^1_s(\CC,{\mathfrak C}^\cdot)$ of
the hyper--cohomology of the complex
$${\mathfrak C}^\cdot:\quad 0\longrightarrow\CO_\CC
{\buildrel{s\cdot}\over\longrightarrow}
\CL\longrightarrow 0$$
of sheaves on $\CC$.
The set of isomorphism classes of infinitesimal deformations of $(\CL,s)$
is isomorphic to ${\mathbb H}^1_s(\CC,{\mathfrak C}^\cdot)$, and a
corresponding Kodaira--Spencer map can be defined for any deformation.

Our goal is to show that the similarity between equation \rref{eq:def}
and the second equation in \rref{eq:sfdb} is not only formal, that is, 
we can interpret Eq.s~\rref{eq:def} as the differential equations associated
with a Kodaira--Spencer deformation of the spectral super line bundle together 
with its meromorphic sections, 
\begin{equation}
  \label{eq:def2}
  \left\{\begin{array}{l}
g_{jk}^{-1}\de_tf_j=(\de_t+g_{jk}^{-1}\de_tg_{jk})f_k
\\
g_{jk}^{-1}\de_\varepsilon f_j=(\de_\varepsilon+g_{jk}^{-1}\de_\varepsilon g_{jk})f_k
\end{array}\right..
\end{equation}
To achieve this we have first of all to construct a suitable family $\CL_\CX$
of line bundles on a $\La$--super--curve $\CC$ and then to interpret the
Fa\`a di Bruno polynomials as local representatives of sections of
$\CL_\CX$. These two steps will be taken in the next two Sections.

%%%%%%%%%%%%%%%%%%%%%%%%%%%%%%%%%%%%%%%%%%%%%%%%%%%%%%%%%%%%%%%%%%%%%%
%
%                U N I V E R S A L   D I V I S O R
%
%%%%%%%%%%%%%%%%%%%%%%%%%%%%%%%%%%%%%%%%%%%%%%%%%%%%%%%%%%%%%%%%%%%%%%

\section{The universal relative positive super divisor}\label{sect:3}

From now on we assume that $\CC$ is a smooth super curve over $\La$.
Since the points of the super universal Grassmannian associated with a
solution of the Jacobian super KP hierarchy belong to the component of
index $0\vert 0$ we must require $\CC$ to be a generic SKP curve and $\CL$ to
have degree $g$ equal to the genus of $\CC$.
\begindef{\bf(SKP curve \cite{BerRab})}
A $\La$--super--curve $\CC=(C,\CO_\CC)$ is called an
{\em SKP curve} if its split structure sheaf
$\CO_\CC^\spt:=\CO_\CC\otimes_\La\La/{\mathfrak m}$
is of the form $\CO_\CC^\rd\vert\CS$, where $\mathfrak m$
is the maximal ideal of nilpotent elements of $\La$, $\CS$ is an
invertible $\CO_\CC^\rd$-module (a ``reduced'' line bundle) of
degree zero and $\cdot\vert\cdot$ denotes a direct sum of free $\La$--modules,
with on the left an evenly generated summand and on the right an odd one.
If $\CS\ne\CO_\CC^{red}$ then $\CC$ is called a {\em generic} SKP curve.
\enddef
Let $\tCC$ be the dual super curve of $\CC$, whose $\La$-points are
the irreducible superdivisors of $\CC$ (see 
Appendix \ref{app:1}).
Constructing a universal family of line bundles $\CL_\CX$ requires the
construction of the 
super Picard scheme of $\CC$ and the corresponding super Poincar\'e sheaf.
However  we can avoid this difficult step, since it suffices
to produce the universal super divisor
$\Da{g}$ of degree $g$.
In analogy to the commutative case, the central object we have to consider 
is the $g$--th symmetric product $S_g\tCC$ of the dual super curve $\tCC$,
since $\tCC$ parameterizes irreducible positive super divisors on $\CC$.
Our discussion will follow closely that of \cite{DPHRSS}, the only novelty
being that we have to work over $\La$, instead over $\BC$. 

Let $\CC^g:=\CC\times_{\spec{\La}}\cdots\times_{\spec{\La}}\CC$ be the 
$g$--fold fibred product of $\CC$ with itself over $\spec{\La}$.
The  symmetric group $\Sigma_g$ of degree $g$ acts on $\CC^g$ by
$$\begin{array}{cccc}
\Sigma_g\ni\sigma: & C^g & \longrightarrow & C^g \\
\\
& (x_1,\cdots,x_g) & \mapsto & (x_{\sigma(1)},\cdots,x_{\sigma(g)})
\end{array}$$
and
\begin{equation}\label{eq:symm}
\begin{array}{cccc}
\sigma: & \CO_\CC^{\otimes_\La g} & \longrightarrow & 
                      \CO_\CC^{\otimes_\La g} \\
\\
& f_1\otimes_\La\cdots\otimes_\La f_g & \mapsto &
\Big(\prod_{\col{j<k}{\sigma(j)>\sigma(k)}}
(-1)^{\bar{f}_{\sigma(j)}\bar{f}_{\sigma(k)}}\Big)
                      f_{\sigma(1)}\cdots f_{\sigma(g)},
\end{array}
\end{equation}
where $C$ is the reduced curve associated with $\CC$.
We define the $g$--th symmetric product of $\CC$ to be the ringed space
$$S_g\CC:=\left(C^g/\Sigma_g,{(\CO^{\otimes_\La g})}^{\Sigma_g}\right),$$
whose structure sheaf is the graded sheaf of invariants of
$\CO^{\otimes_\La g}$. Notice that, since $\sigma$ is an even map (i.e.
it preserves degrees), the action above is the same as that in eq. (1) of
\cite{DPHRSS}. The form given above makes the proof of the following
proposition quite immediate.

\begin{proposition}\label{prop:symm}
The super space $S_g\CC$ is a supermanifold over $\spec{\La}$ of dimension 
$g\vert g$. \end{proposition}
\proof
It is well known that $S_gC:=C^g/\Sigma_g$ is a smooth scheme, so we have to
show that locally $\CO_{S_g\CC}$ is isomorphic to
$\CO_{S_gC}\otimes\La[\theta_1,\ldots,\theta_g]$.
Obviously $\La\subset\CO_{S_g\CC}$.
By definition there exists an open covering $\{U_j\}_{j\in J}$ of $\CC$
such that $\CO_\CC(U_j)\simeq\La\otimes_\BC\CO_C(U_j)[\theta_j]$.
Let $p:C^g\rightarrow S_gC$ be the natural projection of ordinary schemes.
One has only to prove that if $V$ is an open affine subscheme of $S_gC$
such that $\CO_{\CC^g}(U)$ ($U=p^{-1}V$) is isomorphic to 
$\La\otimes_\BC (\CO_{C}(U)[\theta])^{\otimes_\BC g}$, then 
$\CO_{S_g\CC}(V)\simeq \CO_{S_gC}(V)\otimes \La
[\varsigma_1,\ldots,\varsigma_g]$, for suitable odd coordinates $\varsigma_1,\ldots,\varsigma_g$.
Now, $\sigma\in\Sigma_g$ acts as the identity on the first factor $\La$:
in fact we have
\begin{eqnarray*}
&& \sigma(\la_1f_1\otimes\cdots\otimes\la_gf_g) =
     \sigma\left(\prod_{j<k}(-1)^{\bar{f}_j\bar{\la}_k}(\la_1\cdots\la_g)
     f_1\otimes\cdots\otimes f_g\right) \\
&& \phantom{1234} =
    \left(\prod_{\col{l<m}{\sigma(l)>\sigma(m)}}
      (-1)^{(\bar{f}_{\sigma(l)}+\bar{\la}_{\sigma(l)})
            (\bar{f}_{\sigma(m)}+\bar{\la}_{\sigma(m)})}\right)
      \la_{\sigma(1)}f_{\sigma(1)}\otimes\cdots\otimes
            \la_{\sigma(g)}f_{\sigma(g)} \\
&& \phantom{1234} =
     \left(\prod_{\col{l<m}{\sigma(l)>\sigma(m)}}
      (-1)^{\bar{f}_{\sigma(l)}\bar{\la}_{\sigma(m)}+
            \bar{f}_{\sigma(m)}\bar{\la}_{\sigma(l)}+
            \bar{f}_{\sigma(l)}\bar{f}_{\sigma(m)}}\right)
      \left(\prod_{j<k}(-1)^{\bar{f}_{\sigma(j)}\bar{\la}_{\sigma(k)}}\right)
      \times \\
&& \\
&& \phantom{1234567890}
      \la_1\cdots\la_gf_{\sigma(1)}\otimes\cdots\otimes f_{\sigma(g)} \\
&& \\
&& \phantom{1234} =
     \left(\prod_{\col{l<m}{\sigma(l)>\sigma(m)}}
      (-1)^{\bar{f}_{\sigma(l)}\bar{\la}_{\sigma(m)}+
            \bar{f}_{\sigma(m)}\bar{\la}_{\sigma(l)}}\right)
      \left(\prod_{j<k}(-1)^{\bar{f}_{\sigma(j)}\bar{\la}_{\sigma(k)}}\right)
      \times \\
&& \\
&& \phantom{1234567890}
      \la_1\cdots\la_g\sigma(f_1\otimes\cdots\otimes f_g)
\end{eqnarray*}
and since
$$\left(\prod_{\col{l<m}{\sigma(l)>\sigma(m)}}
        (-1)^{\bar{f}_{\sigma(l)}\bar{\la}_{\sigma(m)}+
            \bar{f}_{\sigma(m)}\bar{\la}_{\sigma(l)}}\right)
      \left(\prod_{j<k}(-1)^{\bar{f}_{\sigma(j)}\bar{\la}_{\sigma(k)}}\right)=
\prod_{j<k}(-1)^{\bar{f}_j\bar{\la}_k}$$
we get
$\sigma(\la_1\cdots\la_gf_1\otimes\cdots\otimes f_g)=
\la_1\cdots\la_g\sigma(f_1\otimes\cdots\otimes f_g)$.
Therefore, it remains only to apply Theorem 1 of \cite{DPHRSS}.
If $(z,\theta)$ are graded local coordinates of $\CC$ then a system of 
graded local coordinates for $S_g\CC$ is given by 
$(s_1,\ldots,s_g,\varsigma_1,\ldots,\varsigma_g)$, where $(s_1,\ldots,s_g)$
are the (even) symmetric functions of
$z_j=1\otimes\cdots\otimes z\otimes\cdots\otimes 1$ (with $z$ in
the $j$--th position), $1\le j\le g$ and
$(\varsigma_1,\ldots,\varsigma_g)$ are the odd symmetric functions 
defined by $\varsigma_j:=\sum_{k=1}^g\theta_k\tilde{s}^{(k)}_{j-1}$,
where $\theta_k:=1\otimes\cdots\otimes\theta\otimes\cdots\otimes 1$ and
 $\tilde{s}_j^{(k)}$ is the $j$--th symmetric function of
$z_1,\ldots,z_{k-1},z_{k+1},\ldots,z_g$.
\endpf
To exploit this construction we give the following definition.
\begindef
Let $\CX=(X,\CO_\CX)$ be a super scheme over $\spec{\La}$.
A {\em positive relative super divisor of degree $g$} of
$\CC\times_{\spec{\La}}\CX\to\CX$ is a closed sub--super--scheme $\CZ$ of
$\CC\times_{\spec{\La}}\CX$ of codimension $1\vert 0$, defined by a
homogeneous locally principal ideal $\CJ$ of
$\CO_{\CC\times_{\spec{\La}}\CX}$, such that $\CO_\CZ$ is a locally
free $\CO_\CX$--module of rank $g\vert 0$ and its reduction (modulo
nilpotents) $Z$ is a positive relative divisor of degree $g$ of
$C\times X\to X$.
\enddef
By definition, then, $\CZ$ is locally defined by an equation of type
$$f=z^g-(a_1+\theta\alpha_1)z^{g-1}+\cdots+(-1)^g(a_g+\theta\alpha_g)=0,$$
where $f$ is the local generator of $\CJ$ and the $a_j$'s (respectively the
$\alpha_j$'s) are even (respectively odd) local functions on $\CX$.
Our aim is to show that the symmetric product $S_g\tCC$ is the parameter
space for the universal relative super divisor of degree $g$, $\Da{g}$, of
$\CC$.
The universal relative super divisor of degree $1$ is simply the
sub--super--scheme $\Da{1}$ of $\CC\times_{\spec{\La}}\tCC$ locally
defined by the equation
$$z\otimes_\La 1-1\otimes_\La\tilde{z}-\theta\otimes_\La\tilde{\rho}=0,$$
which we will write more compactly as $z-\tilde{z}-\theta\tilde{\rho}=0$,
where $(z,\theta)$ are local coordinates of $\CC$ and
$(\tilde{z},\tilde{\rho})$ are the ''dual`` coordinates given in Appendix
\ref{app:1}, equation \rref{eq:coord}.
Consider now the natural projections
$$\begin{array}{cccc}
\pi_j: & \CC\times_{\spec{\La}}\tCC^g & \longrightarrow &
         \CC\times_{\spec{\La}}\tCC \\
 & (x,\tilde{x}_1,\ldots,\tilde{x}_g) & \mapsto & (x,\tilde{x}_j)
\end{array}$$
and define $\tDe_j:=\pi_j^{-1}(\Da{1})$, $\tDa{g}:=\tDe_1+\cdots+\tDe_g$.
Since the local equation of $\tDa{g}$ is
$$\prod_{j=1}^g(z-\tilde{z}_j-\theta\tilde{\rho}_j)=
z^g-(s_1+\theta\varsigma_1)z^{g-1}+\cdots+(-1)^g(s_g+\theta\varsigma_g)=0,$$
where the $s_j$'s and the $\varsigma_k$'s are the symmetric functions of
the $\tilde{z}_m$'s and $\tilde{\rho}_n$'s we introduced at the end of
the proof of Proposition \ref{prop:symm}, the next lemma holds true.
\begin{lemma}
There exists a unique positive relative super divisor $\Da{g}$ of degree
$g$ of $\CC\times_{\spec{\La}}S_g\tCC\to S_g\tCC$ such that
$\pi^*(\Da{g})=\tDa{g}$, where
$\pi:\CC\times_{\spec{\La}}\tCC^g\to\CC\times_{\spec{\La}}S_g\tCC$
is the natural projection.\hfill$\square$
\end{lemma}
The most important result we need is Theorem 6 of \cite{DPHRSS}, whose proof 
extends to the present situation.
\begin{theorem}\label{th:univ}
The pair $(S_g\tCC,\Da{g})$ represents the functor of relative positive
super divisors of degree $g$ of $\CC$, i.e. the natural map
$$\begin{array}{cccc}
R: & \Hom(\CX,S_g\tCC) & \to & \Div^g_\CX(\CC\times_{\spec{\La}}\CX) \\
 & f & \mapsto & (id\times f)^*\Da{g}
\end{array}$$
is a functorial isomorphism for every $\La$-super scheme $\CX$.\hfill 
$\square$ \end{theorem}

%%%%%%%%%%%%%%%%%%%%%%%%%%%%%%%%%%%%%%%%%%%%%%%%%%%%%%%%%%%%%%%%%%%%%%
%
%        A L G E B R A I C   G E O M E T R I C   S F D B
%
%%%%%%%%%%%%%%%%%%%%%%%%%%%%%%%%%%%%%%%%%%%%%%%%%%%%%%%%%%%%%%%%%%%%%%

\section{The geometric super Fa\`a di Bruno
polynomials}\label{sect:4}

The constructions of the previous Section allow us to define a canonical
family of super line bundles together with an even section.
For simplicity we call $\CX:=S_g\tCC$ and we assume also that $\CC$ is 
a generic  SKP super curve of genus $g$.
Then $\CL_\CX:=\CO_{\CC\times_{\spec{\La}}\CX}(\Da{g})$ is a
$\CX$--family of super line bundles on $\CC$ and $\Da{g}$ defines a
section $\sigma$ of $\CL_\CX$.
If we let $\CL$ be any non--special super line bundle on $\CC$ (i.e.
such that the reduced invertible sheaf $\CL^\rd$ is non--special on $C$)
of degree $g$ and we call $s_\CL$ the unique (up to multiplication by a
complex number) section that generates the even part of $H^0(\CC,\CL)$,
then the divisor $(s_\CL)$ can be thought of as a $\spec{\La}$--family of
positive relative super divisors of degree $g$ and the universality
property of $\Da{g}$ (Theorem \ref{th:univ}) gives a unique map
$f_\CL:\spec{\La}\to\CX$, i.e. a $\La$--point $x$ of $\CX$, such that
$(s_\CL)=(id\times f_\CL)^*\Da{g}$.
In turn, this induces an isomorphism
$\rho_\CL:\CL\to(id\times f_\CL)^*\CL_\CX$ such that $(id\times
f_\CL)^\ast\sigma=\rho_\CL s_\CL$, so we can interpret the triple
$(\CX,\CL_\CX,\sigma)$ as a deformation of $(\CL,s_\CL)$ for any
non--special super line bundle $\CL$ on $\CC$.
Finally, if we put graded super coordinates ${\bf t}=(t_1,\ldots,t_{2g})$
on $\CX$ ($\bar{t}_j=j\bmod 2$) then the cocycle conditions \rref{eq:def}
for the section $\sigma$ as a deformation of
$s_{{\bf t}_0}:=\sigma\vert_{\CC\times\{x({\bf t}_0)\}}$, for any
${\bf t}_0$, become the differential equations
$$g_{jk}^{-1}\de_{t_l}f_j=\de_{t_l}f_k+(g_{jk}^{-1}\de_{t_l}g_{jk})f_k,$$
which are manifestly of the form of \rref{eq:sfdb}.
To accomplish our goal of describing the algebraic geometric super Fa\`a
di Bruno polynomials, we have therefore only to choose a suitable 
open covering of $\CC\times_{\spec{\La}}\CX$ and to appropriately select 
two coordinates $t_{2j}$ and $t_{2k+1}$ and to call them $x$ and 
$\varphi$ respectively.

As before select a non--special super line bundle $\CL$ of degree $g$ on 
$\CC$. Let $p_\infty\in C$ be a reduced point of $C$ such that it is
not Weierstrass for $C$ and the reduced section $s_\CL^\rd$ does not 
vanish at $p_\infty$.
Let $U_0\subset C$ be an open neighbourhood of $p_\infty$ where we can
define graded coordinates $(z,\theta)$ for $\CC$ centered at $p_\infty$
(i.e. $z(p_\infty)=0$) and let $U_1:=C-\{p_\infty\}$.
Then $\{U_0,U_1\}$ is a Stein open covering of $\CC$.
Since $s^\rd_\CL$ does not vanish at $p_\infty$ the section $s_\CL$ gives a local
trivialization $\eta$ of $\CL$ on $U_0$ (suitably restricted).
Then the quintuple $(\CC,D:=(z)\vert_{U_0},(z,\theta),\CL,\eta)$
defines through the super Krichever map a point of $\Gr$.
Finally, let $\CV$ be a Stein open neighbourhood of $(s_\CL)\in\CX$ where
the coordinates $\bf t$ are defined.
The open subsets $\CU_0:=U_0\times_{\spec{\La}}\CV$ and
$\CU_1:=U_1\times_{\spec{\La}}\CV$ define a Stein covering of
$\CC\times_{\spec{\La}}\CV$ over which we can trivialize
$\CL_\CV:=\CL_\CX\vert_{\CC\times_{\spec{\La}}\CV}$.
Restricting $\CV$ if necessary we can assume that $\sigma$ gives a local
trivialization of $\CL_\CV$ over $\CU_0$.

Now we move to the analytic category instead of the algebraic one.
Let $\CN:=\pi^*\CO_\CC(gD)$, where $\pi$ is now the projection of
$\CC\times_{\spec{\La}}\CV$ to $\CC$, and let $\mu$ be the pull back by
$\pi$ of the section of $\CO_\CC(gD)$ which generates the even
part of its module of global sections.
Then $\mu$ gives a local trivialization of $\CN$ over $\CU_1$.
Since $\CL_\CV\otimes\CN^{-1}$ has relative degree $0$ it follows that,
restricting again $\CV$ if necessary, it has a local analytic
trivialization $\nu$ over $\CU_1$ and $\tau=\nu\mu$ gives a trivialization
of $\CL_\CV$ over $U_1$.

Summarizing, we have a trivialization $(\sigma,\tau)$ of $\CL_\CV$ over
$(\CU_0,\CU_1)$, with respect to which $\sigma$ is represented by the
couple of functions $(f_0=1,f_1)$ and the transition function of $\CL_\CV$
is $g_{10}=f_1/f_0$.
Let us define $\SH{k}:=\de_{t_k}\log g_{10}$.
Then these meromorphic functions on $U_0$ satisfy equation \rref{eq:scomm}
and are therefore our candidates for the super currents of the hierarchy.

Observe that it is possible to choose the coordinates $t_k$ in such a way
that (multiplying $\tau$ by the exponential of a suitable meromorphic
function whose poles are only over $\pi^{-1}(p_\infty)$) $\SH{h}$ has the 
correct
asymptotic behaviour \rref{eq:as} (here our coordinate $z$ is the inverse of
the $z$ appearing there).
Notice also  that $\SH{k}\vert_{\CU_0\cap\CU_1}$ represents the class of
$H^1(\CC\times_{\spec{\La}}\CV,\CO_{\CC\times_{\spec{\La}}\CV})$ corresponding
to the deformation of $\CL$ along $t_k$. Since the asymptotic behaviour of
$\SH{1}$ is $\theta+O(z)$ it follows that the first time $t_1$ does not
deform $\CL$ at all. The super Jacobian $\jac(\CC)$ of $\CC$ has dimension
$g\vert g-1$ while $S_g\tCC$ has dimension $g\vert g$ and maps
surjectively to $\pic^g(\CC)\simeq\jac(\CC$), hence there is an odd
direction in $\CV$ which corresponds to trivial deformations of $\CL$, i.e.
there indeed exists a coordinate like $t_1$.

In \cite{FRZ2}, Section 2.3 we have shown that the Fa\`a di Bruno
generator is computed by the formula
$\hh:=\SH{1}|_{t_1+\varphi}+\varphi\SH{2}|_{t_1+\varphi}$.
The cocycle condition \rref{eq:def} can be interpreted also as saying that
$(\de_{t_k}f_0+\SH{k}f_0,\de_{t_k}f_1)$ is a section of
$\CL_\CV(\infty\pi^*D)$ with pole of order $k$ at $\pi^*D$. Thus the super
Fa\`a di Bruno recurrence relation \rref{eq:sfdb} corresponds to
deformation along the non--integrable vector field $\CD$ and the super
Fa\`a di Bruno polynomials $\sh{k}$ are the local representatives on
$\CU_0$ of the meromorphic sections $\sigma^{(k)}$ of $\CL_\CV$ obtained by
iterative deformation of $\sigma^{(0)}:=\sigma$ along $\CD$.
The form of $\hh$ implies also that the $\sigma^{(k)}$'s form a basis
of $H^{0}(\CC\times_{\spec{\La}}\CV,\CL_\CV(\infty\pi^*D))$ over $\CO_\CV$.
We can restate the above discussion in the following proposition.
\begin{proposition}
The super Fa\`a di Bruno recurrence relation is the cocycle condition
for the hypercohomology group describing the deformations of the dynamical
super line bundle $\CL$ on the spectral curve $\CC$ and of its meromorphic
sections which give rise to the super Krichever map.\hfill $\square$ 
\end{proposition}
We end by remarking that equation \rref{eq:hskp} is an obvious consequence
of the definition of $\hh$ and $\SH{k}$.
\subsection*{Acknowledgments} We thank D. Hern\'andez Ruip\'erez and U. Bruzzo
for useful discussions.  
\appendix

%%%%%%%%%%%%%%%%%%%%%%%%%%%%%%%%%%%%%%%%%%%%%%%%%%%%%%%%%%%%%%%%%%%%%%
%
%                       S U P E R   C U R V E S
%
%%%%%%%%%%%%%%%%%%%%%%%%%%%%%%%%%%%%%%%%%%%%%%%%%%%%%%%%%%%%%%%%%%%%%%

\section{Super curves}\label{app:1}

In this Appendix we recall some facts concerning super curves, referring
to \cite{Gauge} for more details on supergeometry.

Let $\La$ be a Grassmann algebra over $\BC$.
An algebraic super curve over $\La$, also called a {\em $\La$--super--curve}
for brevity, is a proper irreducible superscheme $\CC\to\spec{\La}$ over
$\spec{\La}$ with fibre dimension $1\vert 1$ and whose underlying reduced 
scheme is a proper irreducible algebraic curve over $\BC$.
Throughout this paper we assume $\CC$ to be a supermanifold, so it is
given by a pair $(C,\CO_\CC)$ where $C$ is a topological space and
$\CO_\CC=\CO_{\CC,0}\oplus\CO_{\CC,1}$ is a sheaf of super--commutative 
$\La$--algebras on $C$ such that \begin{itemize}
\item[i.] $(C,\CO_C:=\CO_\CC^\rd=\CO_\CC/\CJ_\CC)$ is a smooth 
irreducible proper algebraic curve over $\BC$, where $\CJ_\CC$ is the ideal
sheaf $\CO_{\CC,1}+\CO_{\CC,1}^2$,
\item[ii.] there exists an open covering $\{U_j\}_{j\in J}$ of $C$ and
odd elements $\theta_j\in\CO_\CC(U_j)$ such that
$$\CO_\CC(U_j)\simeq\CO_C(U_j)\otimes_\BC\La[\theta_j].$$
\end{itemize}
A $\La$--point of $\CC$ is a map $\spec{\La}\to\CC$ whose composition with
the projection $\CC\to\spec{\La}$ is the identity morphism.

An invertible sheaf $\CL$ on $\CC$ is a locally free evenly generated 
$\CO_\CC$--module of rank $1\vert 0$; it is the sheaf of sections of a 
super line bundle that, abusing notations, we still call $\CL$.
We can find a suitable open covering $\{U_j\}_{i\in J}$ of $\CC$ over the
elements of which $\CL$ is trivial. Then the super line bundle is completely
described in terms of its (even invertible) transition functions
$g_{jk}\in\Gamma(U_j\cap U_k,\CO_{\CC,0}^\times)$ satisfying the usual
cocycle conditions.
The set of isomorphism classes of super line bundles on $\CC$ is therefore
$H^1(\CC,\CO_{\CC,0}^\times)$ and tensor product of invertible sheaves 
(or, equivalently, multiplication in $\CO_{\CC,0}^\times$) gives it a 
group structure under which it is called the Picard group $\pic(\CC)$ of 
$\CC$.

Another way to describe an invertible sheaf is by means of super (Cartier)
divisors.
A super divisor on $\CC$ is a collection $D:=\{(U_j,f_j)\}_{j\in J}$ of even
non--zero rational functions $f_j$ defined, up to even invertible regular 
functions, on the open subsets $U_j$ of a covering of $\CC$, and agreeing
in the intersections $U_j\cap U_k$ up to an element of 
$\CO_{\CC,0}^\times(U_j\cap U_k)$: i.e., $D$ is a section of 
$Rat_{\CC,0}^\times/\CO_{\CC,0}^\times$, where $Rat_\CC$ is the sheaf of 
rational functions on $\CC$.
With the super divisor $D$ one associates the invertible subsheaf
$\CO_\CC(D)\subset Rat_\CC$ whose local sections over $U_j$ span the module
$f_j^{-1}\CO_\CC(U_j)$ and the transition functions of the corresponding 
super line bundle are $g_{jk}=f_jf_k^{-1}$.
We have the exact sequence
$$0\to\CO_{\CC,0}^\times\to Rat_{\CC,0}^\times\to
Rat_{\CC,0}^\times/\CO_{\CC,0}^\times\to 0$$
and a super divisor $D$ is called principal if it is the image of a 
global non--zero even rational function $f$, in which case we write $D=(f)$.
Of course, the invertible sheaf associated with a principal divisor is 
trivial and vice versa.
$D$ is called effective (or positive) if $f_j$ is regular for every $j$, and
irreducible if $f_j=z_j-\tilde{z}_j-\theta_j\tilde{\theta}_j$, where
$\tilde{z}_j,\tilde{\theta}_j\in\La$.

A useful concept associated with irreducible super divisors is the dual 
super curve $\tCC$ of $\CC$, which we briefly review (see 
\cite{DRS,BerRab} for more details).
Let $\uCC=(C,\CO_\uCC)$ be the $N=2$ super curve whose reduced algebraic
curve is again $C$ and whose structure sheaf is the only super conformal
extension of $Ber_\CC$ by $\CO_\CC$
\begin{equation}\label{eq:2}
0\to\CO_\CC\to\CO_\uCC\to Ber_\CC\to 0.
\end{equation}
Here $Ber_\CC$ is the dualizing sheaf of $\CC$, whose transition functions
$g_{jk}$ are the Berezinians of the (super) Jacobian matrices of the 
coordinate transformations between $U_j$ and $U_k$, and the super 
conformal property means that the local form
$\omega_j:=\di z_j-\di\theta_j\rho_j$ is globally 
defined up to a scalar factor, where $(z_j,\theta_j,\rho_j)$ are graded 
local coordinates on $\uCC$ adapted to $\CC$ (i.e. $(z_j,\theta_j)$ are
coordinates on $\CC$).
The kernel of $\omega_j$ is generated by $\CD_j:=\de_{\rho_j}$ and
$\tilde{\CD}_j:=\de_{\theta_j}+\rho_j\de_{z_j}$ and one can easily 
convince himself that $\CD_j$ represents locally the map
$\CO_\uCC\to Ber_\CC$, thus the structure sheaf $\CO_\CC$ of $\CC$ is
the kernel of $\CD$.

Introducing the new coordinates
\begin{equation}\label{eq:coord}
\tilde{z}_j:=z_j-\theta_j\rho_j,\qquad
\tilde{\theta}_j:=\theta_j,\qquad
\tilde{\rho}_j:=\rho_j
\end{equation}
on $\uCC$, the two operators above become
$\CD_j=\de_{\tilde{\rho}_j}+\tilde{\theta}_j\de_{\tilde{z}_j}$ and
$\tilde{\CD}_j=\de_{\tilde{\theta}_j}$ respectively, so the kernel of
$\tilde{\CD}_j$ consists of functions of $\tilde{z}_j$ and $\tilde{\rho}_j$.
One shows that this makes sense globally obtaining therefore a new
exact sequence
$$0\to\CO_\tCC\to\CO_\uCC{\buildrel \tilde{\CD} \over \to}\CQ\to 0,$$
where $\CO_\tCC$ is the structure sheaf of a $1\vert 1$
$\La$--super--curve
$\tCC$ which is called the dual super curve of $\CC$, moreover
$\CQ\simeq Ber_\tCC$ and $\tilde{\tCC}\simeq\CC$, which explains the 
terminology.
The interesting fact is that the $\La$--points of $\tCC$ 
correspond to the irreducible divisors of $\CC$.

%%%%%%%%%%%%%%%%%%%%%%%%%%%%%%%%%%%%%%%%%%%%%%%%%%%%%%%%%%%%%%%%%%%%%%
%
%                       B I B L I O G R A P H Y
%
%%%%%%%%%%%%%%%%%%%%%%%%%%%%%%%%%%%%%%%%%%%%%%%%%%%%%%%%%%%%%%%%%%%%%%

\end{document}